# The Expansion of the Universe is Faster than Expected

*Adam G. Riess*

*Johns Hopkins University and the Space Telescope Science Institute*

**Key advances**

- The local or late Universe measurement of the Hubble constant improved from 10% uncertainty twenty years ago to under 2% by the end of 2019.
- In 2019, multiple independent teams presented measurements with different methods and different calibrations to produce consistent results.
- These late Universe estimations disagree at $4\sigma$ to $6\sigma$ with predictions made from the Cosmic Microwave Background in conjunction with the standard cosmological model, a disagreement that is hard to explain or ignore.

The present rate of the expansion of our Universe, the Hubble constant, can be predicted from the cosmological model using measurements of the early Universe, or more directly measured from the late Universe. But as these measurements improved, a surprising disagreement between the two appeared. In 2019, a number of independent measurements of the late Universe using different methods and data provided consistent results making the discrepancy with the early Universe predictions increasingly hard to ignore.

The goal of modern cosmology is to explain the evolution of the Universe from its inception to the present time using our limited understanding of its composition and physical laws. This is even harder than it sounds! The first difficulty materialized in 1929 when initial estimates of the present expansion rate — known as the Hubble constant or $H_0$ — rewound to the Big Bang singularity implied the Universe was younger than the age estimated for the Earth and Sun. In retrospect, both figures were well off the mark, but there has been tremendous progress in the meanwhile. The measurement of $H_0$ improved from 10% uncertainty at the start of the 2000s to under 2% by 2019. In the last few years reduced uncertainties from both the Cosmic Microwave Background (CMB) and local Universe measurements have revealed an underlying discrepancy that is growing harder to ignore.

Scientists now have a "standard model of cosmology, called ΛCDM (Lambda Cold Dark Matter), economically crafted from six free parameters and a number of well-tested ansatzes. The model characterizes a wide-range of phenomena including the accelerating expansion, structure formation, primordial nucleosynthesis, flat geometry of space-time, fluctuations of the Big Bang afterglow and the first combination of baryons into atoms. Remarkably, dark components (matter and energy) account for 95% of the Universe, as described by ΛCDM, their presence robustly inferred from their gravitational effects. Yet despite the success in better understanding our Universe confirmed by a wealth of precise measurements, in the last few years there has been growing evidence that the expansion of the Universe is still exceeding our predictions.

Observations of the CMB —the afterglow of the Big Bang— from the ESA Planck Satellite provide the best present calibration of the parameters in ΛCDM which are used together to provide the most precise estimate of two contemporary quantities; the present age of the Universe of 13.8 ± 0.02 billion years and today's Hubble constant of $H_0$=67.4 ± 0.5 km s$^{-1}$ Mpc$^{-1}$. The use of other measurements from the primordial epoch of the Universe yield a very similar figure. But this estimation from the early Universe is based on the simplest guesses about the nature of dark matter and dark energy and an uncertain roster of relativistic particles (such as neutrinos). A powerful, end-to-end test of ΛCDM and these assumptions is to measure the Hubble constant in the local, or late Universe to a comparable, 1% precision.

The best-established local method is to build a `distance ladder' using simple geometry to calibrate the luminosities of specific star types, pulsating (Cepheid variables) and exploding (Type Ia supernovae or SN Ia), which can be seen at great distances where their redshifts measure the cosmic expansion. A ladder is necessary because the most luminous tools, SN Ia, are too rare to be seen within the range of techniques such as parallax that are purely geometric. The trigonometric parallax can be measured to any visible object, but usefully within only a fraction of the Milky Way. Discovered by Henrietta Leavitt Swan in 1912, Cepheids are supergiant stars whose period of variation strongly correlates with their luminosities. They reach luminosities of 100,000 times that of the Sun which makes them visible with the Hubble Space Telescope in most galaxies to a distance of 40 Mpc. Supernovae reach 10 billion solar luminosities, but only occur once a century in a typical galaxy. Two decades of measurements with these tools have consistently yielded values for $H_0$ in the low 70s.

Then the SH0ES (Supernovae $H_0$ for the Equation of State) project advanced this method by expanding the sample of high quality calibrations of SN Ia by Cepheids increasing the number of independent geometric calibrations of Cepheids and measuring the fluxes of all Cepheids along the distance ladder with the same instrument to negate calibration errors [1]. Improved geometric distance estimates to the Large Magellanic Cloud using detached eclipsing binaries [2] and to galaxy NGC 4258 using water masers [3] have greatly aided this work. The best value of the SH0ES project $H_0=73.5 \pm 1.4$ km s$^{-1}$ Mpc$^{-1}$ is in 4.2σ tension with the early Universe prediction.

An independent, local measurement of $H_0$ comes from the H0LiCOW ($H_0$ Lenses in COSMOGRAIL's Wellspring) team which has been measuring the time delays between multiple images of background quasars to constrain the different image path lengths caused by the strong gravitational lensing from a foreground galaxy. Six such systems have been measured to yield $H_0=73.3\pm1.8$ km s$^{-1}$ Mpc$^{-1}$ [4] with a seventh from a different team, STRIDES (STRong-lensing Insights into Dark Energy Survey), yielding $H_0=74.2\pm3.0$ km s$^{-1}$ Mpc$^{-1}$ [5]. Lensing analyses have markedly improved in the last two decades and these have demonstrated internal consistency from systems with different numbers of quasar images, mean time delays, telescopes and methods used to estimate the lens mass.

Alternative distance ladders have recently been constructed substituting other types of stars in the role usually played by Cepheids. The Tip of the Red Giant Branch (TRGB) is the peak brightness reached by red giant stars after they stop fusing hydrogen and begin fusing helium in their core, a valuable discontinuity which can be observed in SN Ia hosts at distances up to ~20 Mpc with the Hubble Space Telescope (HST). Because the TRGB is not a type of star, but rather a feature of the distribution of thousands, its luminosity cannot be easily calibrated with the limited quality of parallax measurements in the Milky Way, nor readily seen directly with HST in our nearest neighbour, the Large Magellanic Cloud limiting the precision of its calibration. Nevertheless, recent results from TRGB have yielded 69.8±1.9 km s$^{-1}$ Mpc$^{-1}$ [6] and 72.4±1.9 km s$^{-1}$ Mpc$^{-1}$ [7] with the primary difference between these resulting from different estimates of the extinction of TRGB by dust in the Large Magellanic Cloud and of the calibration between HST and ground-based photometry.

Miras are variable, red giant stars recently pressed into service to measure $H_0$ as a check on both Cepheids and TRGB and yield $H_0=73.3\pm3.9$ km s$^{-1}$ Mpc$^{-1}$ using new HST observations and the previously discussed geometric distances [8]. Two other updates to the local measurement of $H_0$ come from water masers (sources of microwave stimulated emission) in four galaxies at great distances ($H_0=74.8\pm3.1$ km s$^{-1}$ Mpc$^{-1}$) and the use of a method called Surface Brightness Fluctuations ($H_0=76.5\pm4.0$ km s$^{-1}$ Mpc$^{-1}$) [9].

Eleven unique averages provide a comprehensive picture of these recent $H_0$ measurements (see Supplementary Information). These were constructed to each exclude a different method (Cepheids, TRGB, Miras, SNe Ia, lensing) or geometric calibration (parallaxes, DEBs, masers in NGC 4258) or team to allow for "peremptory challenges" to the results while using only measurements without overlapping data. They range from 72.8±1.1 km s$^{-1}$ Mpc$^{-1}$ to 74.3±1.0 km s$^{-1}$ Mpc$^{-1}$ and 4.5σ to 6.3σ tension with the Planck estimate. Indeed, it is telling that all recent local measurements exceed the early Universe prediction. Thus, it is hard to avoid the conclusion that the tension with the early Universe prediction is both highly significant and not easily attributed to an error in any one tool, team, or method. New measurements from the local Universe using gravitational lensing and from the early Universe using ground-based CMB polarization are highly anticipated to weigh in over the next few years and may provide new insights.

If the Universe fails this crucial end-to-end test (it surely hasn't yet passed), what might this tell us? It is tempting to think we may be seeing evidence of some `new physics' in the cosmos. Indeed, a large number of theoretical solutions have been proposed and are reviewed in Ref. [10]. For example, if we lived near the middle of a vast and deep void in the large-scale structure of the Universe this could cause excessive, local expansion. However, the odds of a void this large occurring by chance is incredibly low. Calculations show it exceeds 10σ [11] and is also strongly ruled out empirically by the lack of evidence of any end to the void from SNe Ia at greater distances [12]. Dark energy with an equation of state lower than vacuum energy could produce stronger acceleration and explain the discrepancy but this possibility is disfavoured by other intermediate-redshift measurements.

Greater success in explaining the $H_0$ measurement discrepancies has been achieved by altering the composition of the Universe shortly before the emergence of the CMB. An additional component in ΛCDM, such as a new neutrino or scalar field (the latter called Early Dark Energy or EDE), could have increased the early expansion, decreased the sound horizon of primordial fluctuations and raised the predicted value of $H_0$ depending on the approach used to 70-73 km s$^{-1}$ Mpc$^{-1}$ [10] in plausible agreement with the local value. New particles tend to create new conflicts with the CMB whereas EDE is claimed to improve agreement with the CMB. A criticism of EDE is that its scales must be finely-tuned, though the same may be said of the other two episodes of dark energy (inflation and present acceleration). This raises the questions of whether apparent episodes of such anomalous expansion are common or even related. Further, if the true expansion rate is the higher, local value the Universe may actually be up to a billion years younger than expected. More work on the theoretical side and new data are badly needed before we may hope to reach the long sought end-to-end understanding of the Universe.

e-mail: *ariess@stsci.edu*

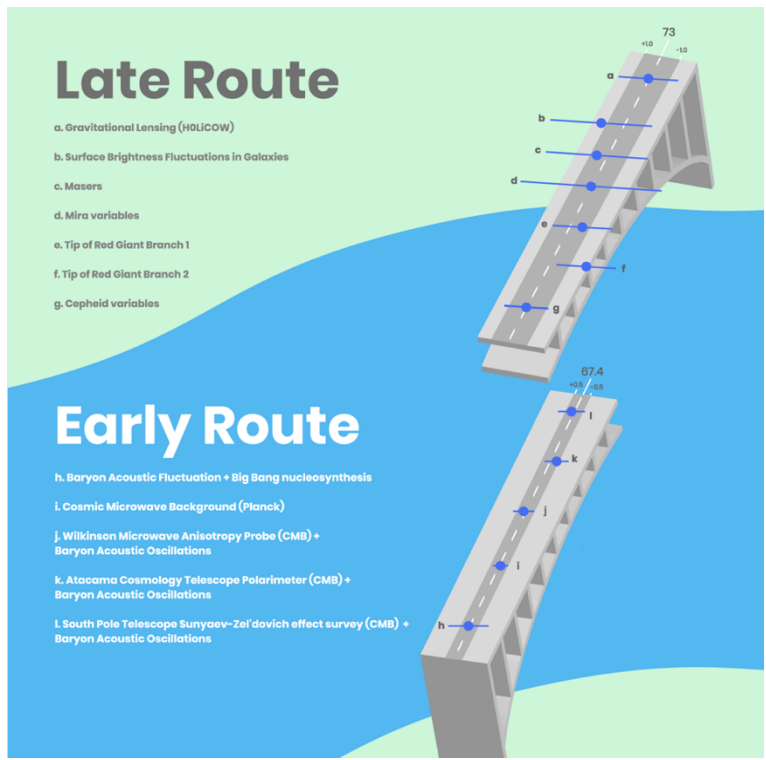

Figure. 1 |. Disagreement between the Hubble constant prediction from the cosmological model using measurements of the early Universe, and the more direct measurements from the late Universe. The discrepancy between the early and late Universe central values represented by the dashed lines at 67.4 km s$^{-1}$ Mpc$^{-1}$ and 73.0 km s$^{-1}$ Mpc$^{-1}$ is of $4\sigma$ to $6\sigma$. The labels denote different measurements: a Planck stands for the ESA Planck Satellite Cosmic Microwave Background observations; b BBN+BAO refers to Big Bang nucleosynthesis and Baryon Acoustic Oscillations (BAO); c SH0ES refers to the SH0ES (Supernovae $H_0$ for the Equation of State) project measurements; d and e refer to the H0LiCOW ($H_0$ Lenses in COSMOGRAIL's Wellspring) team and STRIDES (STRong-lensing Insights into Dark Energy Survey) collaboration estimates, respectively; f and g refer to TRGB (Tip of the Red Giant Branch) estimates; h refers to the estimation using Miras stars; i refers to the estimation using water masers; j SBF refers to the surface brightness fluctuation method; WMAP+BAO refers to measurements using a combination of data from Wilkinson Microwave Anisotropy Probe (WMAP) and BAO; ACTPol+BAO refers to measurements using a combination of data from the Atacama Cosmology Telescope Polarization camera (ACTPol) and BAO; SPT-SZ+BAO refers to measurements using a combination of data from the South Pole Telescope SZ camera (SPT-SZ) and BAO.

# Supplementary Materials

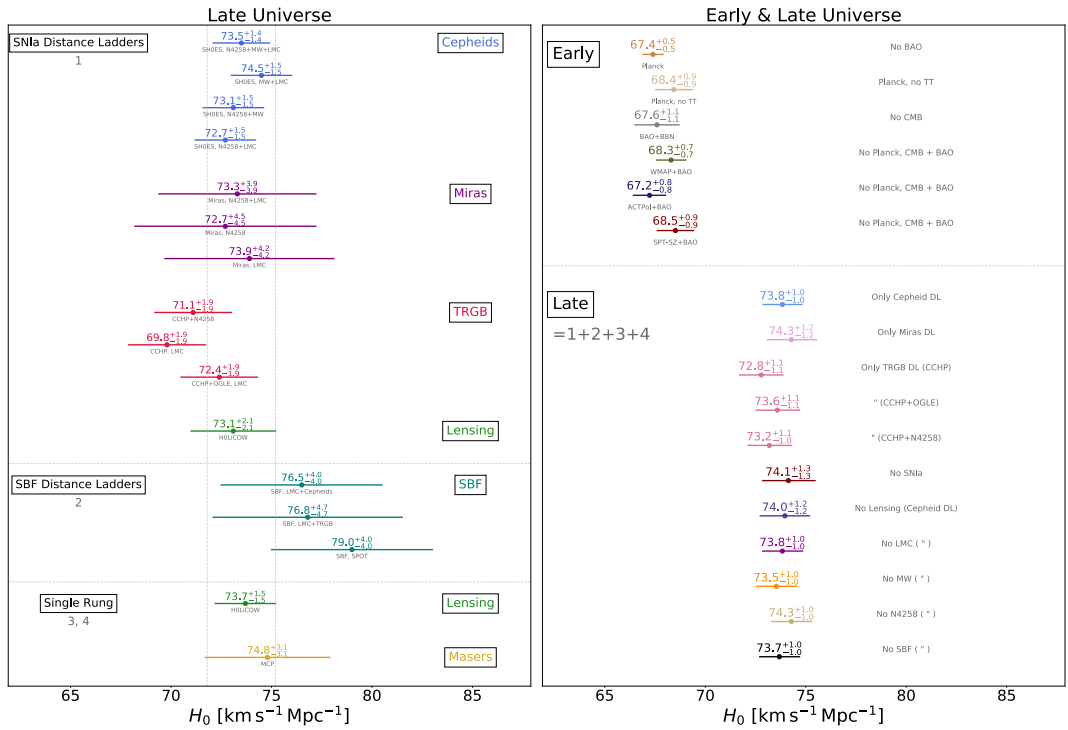

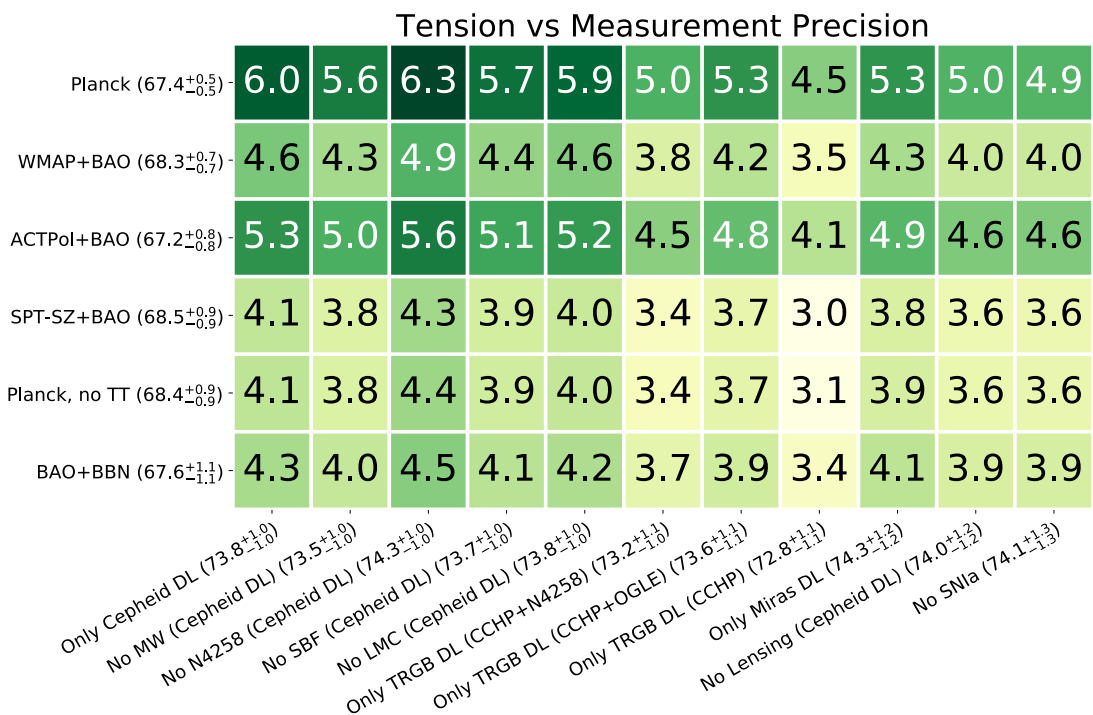